\documentclass[a4paper]{article}

\usepackage{feynmf}

\providecommand{\preprintno}[1]{}

{\def\$#1: #2 ${#2}}

\def\Version/{1.3}
\def\Date/{April 1995}

\def\WOPPER/{\.{WOPPER}}
\def\f77/{\.{FORTRAN-77}}
\def\hepawk/{\.{hepawk}}
\def\hepevt/{\.{/hepevt/}}
\def\HERWIG/{\.{HERWIG}}
\thicklines

\newenvironment{example}[2]{
    \def\examplecaption{#1}
    \def\examplelabel{#2}
    \begin{figure}
    \begin{tabbing}
    \tab\tab\tab\tab\tab\tab\tab\tab\tab\tab\kill
  }{
    \end{tabbing}
    \caption{\examplecaption}
    \label{\examplelabel}
    \end{figure}
  }

\def\.#1{{\tt#1}}                       
\def\,#1{{\rm\it#1\/}}
\def\C{\`\it}                           
\def\tab{\quad\=}

\def\BS/{$\backslash$}

\begin{document}

\title{%
  \WOPPER/, Version \Version/: \\
  A Monte Carlo Event Generator for \\
  $e^+e^- \to (W^+W^-) \to 4f + n\gamma$ \\
  at LEP-II and beyond%
  \thanks{Supported by Bundesministerium f\"ur Forschung und
    Technologie, Germany}}%

\author{%
  Harald Anlauf%
    \thanks{Email: {\tt anlauf@crunch.ikp.physik.th-darmstadt.de}},
  Hans Dieter Dahmen\\
  Universit\"at Siegen\\
  D-57076 Siegen, Germany\\
  \hfil\\
  Angelika Himmler
  Panagiotis Manakos,\\
  Thorsten Ohl%
    \thanks{Email: {\tt Thorsten.Ohl@Physik.TH-Darmstadt.de}}\\
  Technische Hochschule Darmstadt\\
  D-64289 Darmstadt, Germany\\
  \hfil\\
  Thomas Mannel\\
  Institut f\"ur Theoretische Teilchenphysik\\
  D-76128 Karlsruhe, Germany}

\preprintno{IKDA 95/14\\SI 95-3\\hep-ex/9504006}
\date{April 1995}


\maketitle
\begin{abstract}
  We describe the Monte Carlo event generator \WOPPER/ for four
  fermion production through $W$-pairs including resummed leading
  logarithmic QED radiative corrections.
\end{abstract}

\section*{Program Summary:}

\begin{itemize}
\item{} {\bf Title of program:} \WOPPER/, Version \Version/ (\Date/)
\item{} {\bf Program obtainable from:} {\tt
  crunch.ikp.physik.th-darmstadt.de} in the
  directory {\tt /pub/ohl/wopper} (using anonymous Internet ftp)
\item{} {\bf Licensing provisions:} none
\item{} {\bf Programming language used:} \f77/
\item{} {\bf Computer/Operating System:}
  Any with a \f77/ environment
\item{} {\bf Memory required to execute with typical data:} 80k words
\item{} {\bf No.\ of bits in a word:} 32
\item{} {\bf Number of program lines in distributed program, including
    test data, etc.:} $\approx$ 7000 (Including comments)
\item{} {\bf Keywords:} radiative corrections, $W$ pair production, $W$
  decays, multiphoton radiation
\item{} {\bf Nature of physical problem:}  Higher order leading
  logarithmic QED radiative corrections to $W$ pair production and decay
  at high energy $e^+e^-$ colliders, including finite width of the
  $W$'s.
\item{} {\bf Method of solution:} Monte Carlo event generation
\item{} {\bf Restrictions on the complexity of the problem:} The matrix
  elements for the hard subprocess contain only the resonant
  contributions to $e^+e^- \to (W^+ W^-) \to $ 4~fermions in the Born
  approximation.  The final state fermions from the $W$ decays are
  generated at the parton level.
\item{} {\bf Typical running time:} $\approx$ 4 ms/event, depending on
  the energy, cuts and CPU.  The test run takes approximately 50 CPU
  seconds on a 486DX2/66 running Linux 1.2 and 35~CPU seconds on an
  IBM RS/6000-520 running AIX 3.2.5.
\end{itemize}

\newpage

\section{Introduction}
\label{sec:introduction}

The spectacular success of the high energy electron positron colliders
LEP at CERN and SLC at SLAC has confirmed the predictions of the
Standard Model (SM) for the interactions between the gauge bosons and
the fermions even at the level of electroweak radiative corrections.
The high precision of the experimental data allows to put limits on
parameters of yet unobserved particles like
the Higgs boson through the appearance of these particles in weak
loop corrections.  On the other hand, the non-Abelian structure of the
gauge sector of the SM predicts couplings between the electroweak gauge
bosons which have not been tested directly.  The forthcoming upgrade of
LEP to LEP-II \cite{LEP86,Aachen86} with a center of mass energy of up to
$\sqrt{s} \approx 200$~GeV and a future 0.5--1.5~TeV $e^+e^-$ linear
collider~\cite{EE500} (NLC, for short) will in particular make the
trilinear $WW\gamma$ and $WWZ$
couplings directly observable through their contribution to the
production of $W$ pairs.

The main task of LEP-II will be a precise determination of the
mass and the width of the $W$ and the production cross section in the
threshold region~\cite{Delphi92}.  At NLC, a measurement of the total
cross
section, the $W$ angular distributions and in particular of their
longitudinal polarization component will give a handle on possible
anomalous couplings among the electroweak bosons induced by new physics
beyond the SM \cite{BKRS92,HPZH87}, and in turn reveal an insight into
the mechanism of electroweak symmetry breaking.

However, possible new physics is already severely constrained by present
LEP and SLC data, and the effects to be expected at NLC (even more
so at LEP-II) are small.  In order to extract these small
effects, one has to have a precise knowledge of the radiative
corrections within the SM.

The electroweak radiative corrections to the production of on-shell
$W$'s to one-loop order are by now well established \cite{BDS+88,FJZ89}.
The influence of the finite width of the $W$'s has been investigated
in~\cite{MNW86,BBOR93}.  Also, the higher order QED corrections have
been calculated in the leading log approximation (LLA) and their
importance has been emphasized in ref.~\cite{CDMN91}.  Recently an
exhaustive overview of the standard model predictions has been
provided in~\cite{BD94}.

Unfortunately, the experimental reconstruction of the $W$'s and the
determination of their polarization is complicated by the fact that they
may decay either into leptons with an escaping neutrino, or into
hadrons, where the jet energies may be poorly known due to undetected
particles.  In addition, the radiative corrections due to emission of
photons produce a systematic shift of the effective center of mass
energy towards smaller values.  Such effects may best be studied with
the help of a Monte Carlo event generator.

Although quite a few semianalytical calculations of the
corrections mentioned above have been available for some time,
version 1.0 of \WOPPER/ had been the first publicly available
and complete Monte Carlo event generator.  Reference
\cite{Mar91} reported a Monte Carlo generator implementation of the
complete $O(\alpha)$ corrections to the production of on-shell $W$'s.
The matrix elements for single photon bremsstrahlung in the process $e^+
e^- \to (W^+W^-) \to 4$~fermions and their implementation in a Monte
Carlo generator is described in \cite{AW91,Aep91}, but the virtual
corrections are not yet included there.
In the meantime, two new programs have been released~\cite{OFB94} or
are about to be released in the immediate future~\cite{BKP94}.

This paper describes version \Version/ of the Monte Carlo event
generator \WOPPER/.  It is based on the lowest order cross
section for the process $e^+ e^- \to (W^+W^-) \to 4$f and focusses on
QED radiative corrections in the LLA resummed to all orders in $\alpha$
and the effects of finite width of the $W$'s.  The four-momenta of the
exclusive hard photons are generated explicitly and treated with full
kinematics.  The $W$ decays into fermions are implemented at the parton
level, including leading QCD corrections to the $W$ branching fractions.

This write-up is organized as follows: In section~\ref{sec:formalism} we
outline the physics underlying the algorithms implemented in \WOPPER/.
The actual implementation is described in section~\ref{sec:prog-struct}.
The parameters controlling the execution of \WOPPER/ are discussed in
detail in section~\ref{sec:parameters} and the \f77/ interface is
presented in section~\ref{sec:f77}.
Distribution notes, a
listing of all external symbols and the output of a test run can be
found in the appendices.


\section{Theoretical Background and General Features}
\label{sec:general}


\subsection{QED corrections at very high energies}
\label{sec:formalism}

In the structure function formalism \cite{KF85,AM86,BBN89} the
expression for the radiatively corrected cross section reads
\begin{equation}
  \sigma(s) = \int\limits_0^1 dx_+ dx_- \;
  D(x_+,Q^2) D(x_-,Q^2) \; \hat\sigma(x_+ x_- s) \; ,
  \label{eq:factorization}
\end{equation}
where $\hat\sigma$ is the Born level cross section of the hard process,
$D(x,Q^2)$ are the structure functions for initial state radiation, and
$Q^2$ is the factorization scale.

The structure functions $D$ sum the numerically most important leading
logarithmic contributions
\begin{equation}
  \frac{\alpha}{\pi} \log \left(\frac{s}{m_e^2}\right)  \approx 6\%
  \qquad  \mbox{(at LEP-II and NLC energies)}
\end{equation}
to the electromagnetic radiative corrections to all orders.  They
satisfy the QED evolution equation~\cite{KF85}
\begin{eqnarray}
\label{eq:DGLAP}
   Q^2 \frac{\partial}{\partial Q^2} D(x,Q^2)
      & = & \frac{\alpha}{2\pi}
             \int\limits_x^1 \frac{dz}{z} \left[P_{ee}(z)\right]_+
                 D\left(\frac{x}{z},Q^2\right) \\
\label{eq:splitting-function}
  P_{ee}(z) & = & \frac{1+z^2}{1-z}
\end{eqnarray}
with initial condition
\begin{equation}
  D(x,m_e^2) = \delta(1-x) \; .
\end{equation}
The solution to eq.~(\ref{eq:DGLAP}) automatically includes the very
important exponentiation of the soft photon contributions to the
radiative corrections.

An explicitly regularized version of (\ref{eq:DGLAP})
used in the Monte Carlo implementation is given by
\begin{eqnarray}
\label{eq:regularized-DGLAP}
   Q^2 \frac{\partial}{\partial Q^2} D(x,Q^2)
     & = & \frac{\alpha}{2\pi}
              \int\limits_x^{1-\epsilon} \frac{dz}{z} P_{ee}(z)
                 D\left(\frac{x}{z},Q^2 \right) \\
     &   & \mbox{} - \frac{\alpha}{2\pi}
             \left[ \int\limits_0^{1-\epsilon} dz P_{ee}(z) \right]
                  D(x,Q^2). \nonumber
\end{eqnarray}
It is crucial to note that the characteristics of the generated event
sample do not depend on~$\epsilon$, if it is chosen well below the
experimental threshold for the detection of soft photons.  Theoretically,
the cross sections will remain positive for {\em all\/} values of
$\epsilon$, but limited storage for ultra soft photons and limited
floating point range impose a lower limit.

The radiatively corrected cross section (\ref{eq:factorization}) is
implemented in a Monte Carlo event generator by solving the
integro-differential equation~(\ref{eq:regularized-DGLAP}) by iteration
and taking into account the energy loss in the hard cross section.  It
is very similar to algorithms for quark fragmentation in
QCD~\cite{Sjo85}.  As a by-product of the branching algorithm, the
four-momenta of the radiated photons are generated explicitly.

The initial state branching algorithm has already been used in the Monte
Carlo generator {\tt KRONOS} \cite{ADM+92a} and an improved version in
the generator {\tt UNIBAB} \cite{UNIBAB}, where the implementation has
been described in detail.

The momenta of the electron and positron after initial state branching
are then used as the input momenta for the subgenerator of the hard
subprocess described below.


\subsection{Born Cross Section}
\label{sec:born}

In the general case, there are many Feynman diagrams contributing to the
process $ e^+e^- \to $ 4~fermions at high energies, even if one requires
that the quantum numbers of the final state fermions be consistent with
$W$ pair production.  However, the contribution of the individual
Feynman diagrams may be easily estimated by counting the number of
`resonant' propagators, where an intermediate vector boson may come
close to its mass shell.  Using this naive estimate, one finds that the
contribution of these `background diagrams' is suppressed by a factor
$\Gamma_W/M_W \sim $2.5\% for each non-resonant boson propagator, and
may be reduced further by appropriate cuts on the invariant masses of
the reconstructed $W$'s.

In the current version \Version/ of \WOPPER/ only the Feynman diagrams
with two resonating $W$'s contributing to the process $e^- e^+ \to (W^+
W^-) \to f_1 \bar f_2 f_3 \bar f_4$ are taken into account.  In fact, a
full calculation~\cite{BKP94} finds that this approach is numerically
justified for LEP-II energies, unless electrons are in the final state
and no invariant mass cuts are applied.

The implementation of the Born cross section in \WOPPER/ is split into
two steps, namely the production of the virtual (off-shell) $W$'s and
their subsequent decay into fermions including the correlations from the
polarization of the intermediate $W$'s.

This factorization in the resonant approximation can easily be seen as
follows.  Since the top quark is much heavier than the $W$, the virtual
$W$'s will decay only into (almost) massless fermions.  We choose the
Landau gauge.  The couplings of the would-be Goldstone bosons to the
fermions are proportional to the fermion masses and may therefore be
safely neglected.  The numerator of the $W$ propagators is decomposed
into a sum over three physical polarizations.
\begin{equation}
  -g_{\mu\nu} + \frac{k_\mu k_\nu}{k^2} =
  \sum_{\lambda=1}^3 \epsilon_\mu(k,\lambda) \epsilon^*_\nu(k,\lambda)
  \; , \quad
  k \cdot \epsilon(k,\lambda) = 0
  \; , \quad
   \epsilon(k,\lambda) \cdot \epsilon^*(k,\lambda') =
   -\delta_{\lambda\lambda'}
\end{equation}
As in ref.~\cite{HPZH87}, we write the amplitude for
\begin{equation}
  e^-(p_1) e^+(p_2) \to
  W^-(k_-) W^+(k_+) \to
  f_1(q_1) \bar f_2(q_2) f_3(q_3) \bar f_4(q_4)
\end{equation}
as a coherent sum of a production and a decay amplitude for virtual
$W$'s,
\begin{eqnarray}
  \label{eq:full-amplitude}
  {\cal M}  & = &  \sum_{\lambda_+, \lambda_-}
  {\cal M}_{e^+e^- \to W^+ W^-}(p_1,p_2;k_-,\lambda_-;k_+,\lambda_+)
  \\ && \quad \times
  Z(k_-^2)^{-1} \; Z(k_+^2)^{-1} \;
  {\cal M}_{W^- \to f_1 \bar f_2}(\lambda_-; q_1,q_2) \;
  {\cal M}_{W^+ \to f_3 \bar f_4}(\lambda_+; q_3,q_4)
  \nonumber
\end{eqnarray}
where
\begin{equation}
  Z(k^2)^{-1} \; = \;
  \frac{1}{k^2 - M_W^2 + \Sigma_W(k^2)} \; .
\end{equation}
Although we are working in the Born approximation, we shall use the
renormalized $W$ self energy $\Sigma_W(k^2)$ in this expression in order
to obtain a realistic behavior of the amplitude when a $W$ in the
intermediate state comes close to its mass shell.  For LEP-II energies
and in the on-shell renormalization scheme, it is sufficient to take
only the leading linear term of the energy dependence of the imaginary
part of $\Sigma_W$ into account and to neglect the real part.
\begin{equation}
  \mbox{Im } \Sigma_W(k^2) \equiv \sqrt{k^2} \cdot \Gamma_W(k^2)
  \approx \frac{k^2}{M_W^2} \; M_W \Gamma_W \; , \qquad
  \mbox{Re } \Sigma_W(k^2) \approx 0
  \label{eq:sigma}
\end{equation}

\expandafter\ifx\csname fmffile\endcsname\relax\else
\begin{fmffile}{manpics}
\begin{figure}[tb]
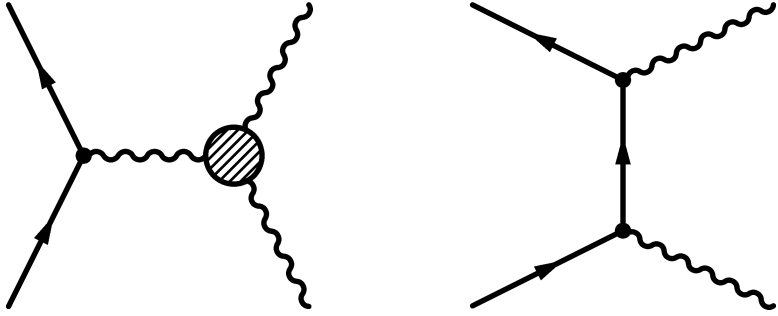

  \begin{center}
    \begin{fmfchar*}(50,40)
      \fmfpen{thick}
      \fmfleftn{i}{2} \fmfrightn{o}{2}
      \fmf{fermion,label=$p^-$,label.side=left}{i1,v1} \fmflabel{$e^-$}{i1}
      \fmf{fermion,label=$p^+$,label.side=left}{v1,i2} \fmflabel{$e^+$}{i2}
      \fmf{photon,label=$\noexpand\gamma,,Z$}{v1,v2}
      \fmf{photon,label=$k^-$,label.side=left}{v2,o1} \fmflabel{$W^-$}{o1}
      \fmf{photon,label=$k^+$,label.side=right}{v2,o2} \fmflabel{$W^+$}{o2}
      \fmfblob{(.15w)}{v2}
      \fmfdot{v1}
    \end{fmfchar*}
    \qquad\quad
    \begin{fmfchar*}(50,40)
      \fmfpen{thick}
      \fmfleftn{i}{2} \fmfrightn{o}{2}
      \fmf{fermion,label=$p^-$,label.side=left}{i1,v1} \fmflabel{$e^-$}{i1}
      \fmf{fermion,label=$\noexpand\nu_e$,label.side=left}{v1,v2}
      \fmf{fermion,label=$p^+$,label.side=left}{v2,i2} \fmflabel{$e^+$}{i2}
      \fmf{photon,label=$k^-$,label.side=left}{v1,o1} \fmflabel{$W^-$}{o1}
      \fmf{photon,label=$k^+$,label.side=right}{v2,o2} \fmflabel{$W^+$}{o2}
      \fmfdotn{v}{2}
    \end{fmfchar*}
  \end{center}
  \caption{The Feynman diagrams contributing to the process $e^+
    e^- \to W^+ W^-$.}
  \label{fig:diagrams}
\end{figure}
\end{fmffile}
\fi

Taking the modulus squared of (\ref{eq:full-amplitude}) and integrating
over the decay angles of the virtual $W$'s, one obtains the well known
resonance formula \cite{MNW86,DS90,Aep91} for the total cross section
for production of virtual $W$ pairs (see figure~\ref{fig:diagrams}).
\begin{equation}
  \sigma_{res} \; = \;
  \int ds_+ ds_- \;
  \frac{\sqrt{s_+} \,\Gamma_W(s_+)}{\pi D(s_+)}
  \frac{\sqrt{s_-} \,\Gamma_W(s_-)}{\pi D(s_-)}
  \sigma_{off}(s; s_+, s_-)
  \label{eq:resonance-formula}
\end{equation}
Here $\sigma_{off}(s; s_+, s_-)$ denotes the off-shell cross section for
the production of two $W$'s, and $\Gamma_W(s_\pm)$ is the effective
decay width of the virtual $W$'s into light fermions according to
eq.~(\ref{eq:sigma}).  The resonance factors are essentially
Breit-Wigner functions, but with energy-dependent width
\begin{equation}
  \frac{1}{D(s_\pm)} \; = \;
  \frac{1}{(s_\pm - M_W^2)^2 + s_\pm \Gamma_W^2(s_\pm)} \; .
\end{equation}
For the actual implementation it is useful to apply the mappings
($\gamma \equiv \Gamma_W/M_W$)
\begin{equation}
  \xi_\pm =
  \arctan \left( \frac{(1+\gamma^2) s_\pm - M_W^2}{\gamma M_W^2} \right)
  \label{eq:mapping}
\end{equation}
to eq.~(\ref{eq:resonance-formula}) in order to get a smooth integrand
suitable for a Monte Carlo rejection algorithm:
\begin{eqnarray}
  \sigma_{res}  & = &
  \left. \int d\xi_+ d\xi_- \;\;
  \frac{1}{\pi} \frac{s_+}{M_W^2} \;
  \frac{1}{\pi} \frac{s_-}{M_W^2} \;
  \sigma_{off}(s; s_+, s_-)
  \right|_{s_\pm = M_W^2 \cdot (1 + \gamma\tan \xi_\pm) / (1+\gamma^2)}
  \nonumber \\  & \equiv &
  \int d\xi_+ d\xi_- \; \tilde\sigma(s; \xi_+, \xi_-)
  \label{eq:resonance-formula-mapped}
\end{eqnarray}

The angular distribution of the virtual $W$'s in the center of mass
frame is generated from the differential cross section
$d\sigma(s;s_+,s_-;\theta)/d\Omega$ by a standard mapping and rejection
algorithm.

\subsection{Coulomb singularity}
\label{sec:coulomb}

Another class of universal corrections, which is important near
threshold, is the so-called Coulomb singularity.  We implement the
correction~\cite{BBD93}
\begin{equation}
  \sigma_{\scriptsize\rm Coulomb}
     = \sigma_{\scriptsize\rm Born} \frac{\alpha\pi}{2\beta}
      \left(1 - \frac{2}{\pi} \arctan\left(
        \frac{\vert\beta_M+\Delta|^2 - \beta^2}%
             {2\beta\mathop{\rm Im}\beta_M}\right)\right)
\end{equation}
to the off shell total cross section.


\subsection{Decays of the virtual $W$'s}
\label{sec:decay}

After the momenta of the virtual $W$'s have been fixed, it is
straightforward to generate the decays of the $W$'s in the center of mass
system.

The measure of the integrations over the $W$-decay angles
is independent of the other
kinematical variables, thus leading to the simple form of
eq.~(\ref{eq:resonance-formula}).
Hence, the decay angle distributions for fixed
virtualities and production angles of the $W$'s are proportional to the
modulus squared of the amplitude~(\ref{eq:full-amplitude}), which is a
sufficiently smooth function of the decay angles and therefore suitable
for a rejection algorithm.

The final state fermions of the decay $W \to f \bar f'$ are simply
chosen with a probability according to the branching fractions obtained
from the tree-level formulas.  Inclusive QCD corrections to the $W$
decays to quarks are taken into account to first order in $\alpha_S$.
The branching fractions for hadronic decays into~$u$, $d$, $c$, $s$,
and~$b$~quarks are given by the corresponding CKM matrix elements,
while decays into~$t$~quarks are assumed to be kinematically
forbidden because of~$|m_t-M_W|\gg\Gamma_W$.
Currently, the fermion masses are only taken into account
kinematically and not in the matrix elements.


\subsection{Hadronization}

If one or both of the $W$'s have decayed into quarks, they can
optionally be hadronized using either the LUND string
model~\cite{JETSET} or the HERWIG cluster model~\cite{MWA+92}.
In the present version it is not possible\footnote{%
  Unless one makes the necessary changes to the \WOPPER/ sources in
  routines \.{wwlund()} and \.{wwhwig()}.}
to study color-rearrangement effects in purely hadronic decays.  Both
$W$~decays are handled separately.

The hadronization model can be switched at run time with the parameter
\.{qcdmc}, which takes the values 0, 1 and 2.  These correspond to no
QCD MC, {\tt JETSET}~\cite{JETSET} and {\tt HERWIG}~\cite{MWA+92}
respectively.


\section{Implementation of \WOPPER/ \Version/}
\label{sec:prog-struct}

Like almost all Monte Carlo event generators, \WOPPER/ is divided into
three parts: initialization, event generation, and termination.  These
are described in this section.  For ease of use, \WOPPER/ comes with two
application interfaces, so that a direct call to the lower level parts
in this section will never be necessary.  These interfaces will be
explained in section~\ref{sec:f77}.


\subsection{Initialization}
\label{sec:initialization}

The initializations in \WOPPER/ are used for computing the value of
variables that will be used frequently during event generation.
Examples are the calculation of electroweak couplings, the maximum of
the off-shell Born cross section and other internal steering parameters
from the input parameters.  This is accomplished by a call to the
subroutine \.{wwinit} after setting the Monte Carlo parameters.  Since
\WOPPER/ does not yet include weak corrections, most of the
initializations performed are quite trivial.

Finally, a standard \hepevt/ initialization record \cite{AKV89} is
written, which may be used by the analysis program.


\subsection{Event Generation}
\label{sec:generation}

The routine \.{wwgen} produces an event on every call.  The four momenta
of all generated particles as well as supplemental information is
written to a standard \hepevt/ event record, where it can be read by
user supplied analyzers.  See section~\ref{sec:hepevt} below for details
on \WOPPER/'s extensive use of \hepevt/.

The first step is the generation of the initial state radiation by the
branching routine \.{wwbini} using the algorithm of
section~\ref{sec:formalism} and of the virtual masses of the $W$'s
occurring in the intermediate state.  According to the cross section
(\ref{eq:resonance-formula-mapped}), the rejection weight is calculated
from the ratio of the actual off-shell cross section to the maximum
determined in the initialization step, and the event is accepted with a
probability according to this weight.

After the effective center of mass energy and the $W$ virtualities have
been fixed, the angular distribution of the virtual $W$'s is generated
in \.{wwgt} using the helicity amplitudes \.{wwxhel} for off-shell $W$
pair production.  Finally, the decay of the intermediate $W$'s into the
final state fermions is accomplished in the subroutine \.{wwgdec}.


\subsection{Termination}

The cross section for the generated events is obtained in the subroutine
\.{wwclos} from the standard formula
\begin{equation}
  \sigma_{tot}(s) =
  \max_{s';\xi_+,\xi_-} \{\tilde\sigma(s';\xi_+,\xi_-)\}
  \cdot \frac{\mbox{\# of successful trials}}{\mbox{total \# of trials}}
\end{equation}
where $\tilde\sigma$ is the off-shell cross section from
eq.~(\ref{eq:resonance-formula-mapped}).  $s'$ varies between 0 and $s$,
and $\xi_\pm$ in the range allowed by the specified cuts.  The
corresponding statistical error from the Monte Carlo integration is
given by
\begin{eqnarray}
  \Delta\sigma_{tot}(s)  & = &
  \max_{s';\xi_+,\xi_-} \{\tilde\sigma(s';\xi_+,\xi_-)\} \cdot \\
  && \sqrt{\frac{(\mbox{total \# of trials} - \mbox{\# of successful
   trials}) \cdot \mbox{\# of successful trials}}
   {(\mbox{total \# of trials})^3}} \nonumber
\end{eqnarray}
The cross section and the error on the cross section are placed into
\hepevt/, where they may be read by the user-supplied analyzer.


\subsection{Additional information in \hepevt/}
\label{sec:hepevt}

Because \WOPPER/ uses the standard \hepevt/ event record internally, not
only stable particles with \.{isthep(i) = 1} will be present.  Adapting
the conventions of the \HERWIG/ Monte Carlo \cite{MWA+92}, we use the
following status codes
\begin{itemize}
  \item{} 101: $e^-$ beam (positive $z$-direction),
  \item{} 102: $e^+$ beam (negative $z$-direction),
  \item{} 103: center of mass system of the collider,
  \item{} 110: $e^+e^-$ hard scattering center of mass system,
  \item{} 111: $e^-$ before hard scattering,
  \item{} 112: $e^+$ before hard scattering,
  \item{} 113: virtual $W^-$ after hard scattering,
  \item{} 114: virtual $W^+$ after hard scattering.
\end{itemize}
However, these entries have {\em no\/} physical significance and should
{\em never\/} be used in any analysis (an exception to this rule are the
beam particles 101 and 102, which are convenient for defining the
reference frame and are used e.g.~by the analyzer \hepawk/ \cite{Ohl92a}
for this purpose).  Only the particles with status code 1 belong to the
final state as predicted by \WOPPER/.

If no hadronization Monte Carlo is active, final state quarks will be
entered as ``stable'' particles with status code 1.


\section{Parameters}
\label{sec:parameters}

The parameters controlling \WOPPER/ version \Version/ are summarized
in tables \ref{tab:wopper-physics-parm} and
\ref{tab:wopper-technical-parm}.  They will be described in the
following subsections. 

\begin{table}
  \begin{minipage}{\textwidth}
  \begin{center}
  \begin{tabular}{@{\vrule width 0pt height 15pt depth 5pt}|c|c|c|}
    \hline\hline
    Variable name   & semantics             & Default value
    \\\hline\hline
    \.{ebeam}       & $e^+$/$e^-$ beam energy & 250 GeV
    \\\hline
    \.{epol}        & longitudinal $e^-$ beam polarization & 0
    \\\hline
    \.{ppol}        & longitudinal $e^+$ beam polarization & 0
    \\\hline
    \.{scheme}      & renormalization scheme  & 1
    \\\hline
    \.{ahpla}       & $1/\alpha_{QED}(4M_W^2)$ & 128
    \\\hline
    \.{ahpla0}      & $1/\alpha_{QED}(0)$   & 137.0359895
    \\\hline
    \.{mass1e}      & $m_{e^\pm}$           &
                                         $0.51099906\cdot 10^{-3}$ GeV
    \\\hline
    \.{mass1w}      & $M_{W^\pm}$           & 80.22  GeV
    \\\hline
    \.{gamm1w}      & $\Gamma_{W}$          &  2.08  GeV
    \\\hline
    \.{mass1z}      & $M_{Z^0}$             & 91.187 GeV
    \\\hline
    \.{gamm1z}      & $\Gamma_{Z}$          &  2.492 GeV
    \\\hline
    \.{gfermi}      & $G_F$                 &  $1.16639\cdot 10^{-5}$ GeV
    \\\hline
    \.{sin2w}       & $\sin^2\theta_W$      &  0.2261
    \\\hline
    \.{alphas}      & $\alpha_{QCD}(M_W^2)$ &  0.12
    \\\hline
    \.{ckmvus}      & $V_{us}=\sin\theta_C$ &  0.2196
    \\\hline
    \.{ckmvcb}      & $V_{cb}$              &  0.0400
    \\\hline
    \.{ckmvub}      & $V_{ub}$              &  0.0032
    \\\hline
  \end{tabular}
  \end{center}
  \end{minipage}
  \caption{Physics parameters controlling \WOPPER/.}
  \label{tab:wopper-physics-parm}
\end{table}

\begin{table}
  \begin{minipage}{\textwidth}
  \begin{center}
  \begin{tabular}{@{\vrule width 0pt height 15pt depth 5pt}|c|c|c|}
    \hline\hline
    Variable name   & semantics             & Default value
    \\\hline\hline
    \.{cutmin}      & minimum $W^\pm$ virtuality & 0 GeV
    \\\hline
    \.{cutmax}      & maximum $W^\pm$ virtuality & $\sqrt{s} = 2 E_{Beam}$
    \\\hline
    \.{nevent}      & Number of events      & 10000
    \\\hline
    \.{cc}          & apply canonical cuts  & .false.
    \\\hline
    \.{cclvl}       & level of canonical cuts  & 0
    \\\hline
    \.{ccver}       & version of canonical cuts  & 1
    \\\hline
    \.{bstyle}      & Key for QED radiation & 1
    \\\hline
    \.{epsiln}      & Internal infrared cutoff & $10^{-5}$
    \\\hline
    \.{coulom}      & Include final state coulomb corrections & {\tt .false.}
    \\\hline
    \.{qcdmc}       & Key for QCD Monte Carlo  & 0
    \\\hline
    \.{rangen}      & Random number generator  & 1
    \\\hline
    \.{rseed}       & Random number seed       & 54217137
    \\\hline
    \.{rlux}        & `Luxury' of {\tt ranlux} & 3
    \\\hline
    \.{errmax}      & maximum error count      & 100
    \\\hline
    \.{verbos}      & verbosity                & 0
    \\\hline
    \.{runid}       & run identification       &
    \\\hline
    \.{stdin}       & standard input           & 5
    \\\hline
    \.{stdout}      & standard output          & 6
    \\\hline
    \.{stderr}      & standard error           & 6
    \\\hline
  \end{tabular}
  \end{center}
  \end{minipage}
  \caption{Technical parameters controlling \WOPPER/.}
  \label{tab:wopper-technical-parm}
\end{table}


\subsection{Electroweak Parameters}

Since the present version of \WOPPER/ does not include an electroweak
library, the electroweak parameters, namely the masses of the
electroweak bosons (\.{mass1w}, \.{mass1z}), their widths (\.{gamm1w},
\.{gamm1z}) and the Weinberg angle (\.{sin2w}) are treated as
independent parameters.  They enter an effective Born cross section and
may be set directly by the user.  As a special case, setting \.{gamm1w}
to 0 reproduces the narrow-width approximation with on-shell $W$'s in
the intermediate state.

The bulk of the non electromagnetic radiative corrections can be
incorporated into the hard cross section by using the running QED
coupling $\alpha_{QED}(4M_W^2) \approx 1/128$ at the $W$ scale.  This value
is. however, {\em not\/} correct for the initial state radiation of
on-shell photons, where $\alpha_{QED}(0) \approx 1/137$ has to be
used.  The inverse of the former value can be changed with \.{ahpla}
and that of the latter with \.{ahpla0}.

The presence of initial state radiation can be toggled using the
parameter \.{bstyle}.  The supported supported values are 0 and 1,
corresponding to no QED radiative corrections and LLA resummed initial
state QED radiative corrections.

Since version 1.3 it is possible to choose canonical input parameters
for benchmarking LEP2 Monte Carlos~\cite{LEP2-YB} by changing the
value of \.{scheme}. The follow values are supported:
\begin{itemize}
  \item{} \.{scheme = 0}: ``free scheme'', all parameters are taken
    from the input and treated as independent parameters.
  \item{} \.{scheme = 1} (default): ``$G_F$ scheme''
    \begin{eqnarray}
      \sin^2\theta_W & = &
        \frac{\pi\alpha_{QED}(4M_W^2)}{\sqrt{2} G_F M_W^2} \\
      \Gamma_W & = & \frac{G_F M_W^3}{\sqrt{8}\pi}
                     \left(3+ \frac{2\alpha_{QCD}}{\pi}\right)
    \end{eqnarray}
  \item{} \.{scheme = 2}: ``$\sin\theta_W^{\textrm{eff.}}$ scheme''
    \begin{eqnarray}
      G_F & = &
        \frac{\pi\alpha_{QED}(4M_W^2)}{\sqrt{2} \sin^2\theta_W M_W^2} \\
      \Gamma_W & = & \frac{G_F M_W^3}{\sqrt{8}\pi}
                     \left(3+ \frac{2\alpha_{QCD}}{\pi}\right)
    \end{eqnarray}
  \item{} \.{scheme = 3}: ``Born scheme'', tree level formulas,
    independent $\sin\theta_W$, $G_F$,
    $\alpha_{QED}(4M_W^2)=\alpha_{QED}(0)$ and
    \begin{eqnarray}
      \Gamma_W & = & \frac{G_F M_W^3}{\sqrt{8}\pi}
                     \left(3+ \frac{2\alpha_{QCD}}{\pi}\right)
    \end{eqnarray}
  \item{} \.{scheme = -1, -2, -3}: same as the positive values,
    except for~$\Gamma_W$ which is not calculated from the standard
    model expression but taken from \.{gamm1w} instead.
\end{itemize}


\subsection{Cuts}

In the present version \Version/ of \WOPPER/, only cuts in the
virtualities of the intermediate $W$'s are implemented in the event
generation:
\begin{itemize}
  \item{} \.{cutmin}: minimum virtuality of the intermediate $W^\pm$s,
  \item{} \.{cutmax}: maximum virtuality of the intermediate $W^\pm$s.
\end{itemize}
The cuts in virtualities have to satisfy the following conditions:
\begin{equation}
  0 \le \mbox{\.{cutmin}} < \mbox{\.{cutmax}} \le \sqrt{s} = 2 E_{Beam}
  \label{eq:cut-conditions}
\end{equation}
A value of 0 for \.{cutmax} will automatically be reset to the available
maximum, namely $2 E_{Beam}$.

Since version 1.3 it is possible to apply the canonical cuts for
benchmarking LEP2 Monte Carlos~\cite{LEP2-YB} to the event record by
setting \.{cc} to \.{.true.}.  These cuts will be reflected in the
caclulated total cross section.


\subsection{Monte Carlo Parameters}

The remaining, more technical Monte Carlo parameters should be almost
self explaining.  Since our branching algorithm automatically includes
soft photon exponentiation, the results will not depend on the value of
the internal infrared cutoff \.{epsiln} (which is measured in units of
the beam energy), provided it is kept {\em well below\/} the
experimental energy resolution.  However, it is not advisable to set it
many orders of magnitude lower than the default value, because this may
result in too high photon multiplicities that will overflow internal
tables.

A note on the random number generators available: the default value~1
of the parameter \.{rangen} corresponds to the standard
\.{RANMAR}~\cite{MZT90}
generator, that has been the generator of choice for quite some time.
Recently, the quality of the random numbers generated by \.{RANMAR}
has been questioned and unwanted correlations have been found, that
caused large systematic errors in solid state physics
simulations~\cite{FLW92}.  A superior variation \.{RANLUX} has been
proposed~\cite{Lue93}, which is however {\em much\/} slower.
Setting \.{rangen} to~2 switches to \.{RANLUX} which can be used at
the 5 ``luxury levels'' 0 to~4.  At the highest ``luxury levels'',
\WOPPER/ will spent more than half of the total computer time in the
random number generator.

However, since event generation involves a lot of decisions that
effectively randomize the subsequences used by themselves, we do not
expect that the correlations in \.{RANMAR} have any significant effect
on the event samples generated by \WOPPER/.  The \.{RANLUX} option has
been added to \WOPPER/ for some experimentation only.  It is left in
only because there is no particular reason for throwing it out again.


\subsection{QCD Parameters}

The parameters for \.{JETSET} should be accessed trough \WOPPER/'s
\.{lugive} command, which is translated directly to \.{JETSET}'s
\.{LUGIVE()} subroutine.  See the \.{JETSET} manual~\cite{JETSET} for
a comprehensive description of the available parameters and their
effects.

Since \.{HERWIG} does not sport the equivalent of the \.{LUGIVE()}
routine, its parameters have to be accessed through the standard
\WOPPER/ access mechanisms.  Tables~\ref{tab:wopper-herwig-parm}
and~\ref{tab:wopper-technical-herwig-parm} provide a list of the
available \.{HERWIG} parameters and the names under whith they are known
to \WOPPER/.  See the \.{HERWIG} manual~\cite{MWA+92} for a
comprehensive description of the effects of these parameters.

\begin{table}
  \begin{minipage}{\textwidth}
  \begin{center}
  \begin{tabular}{@{\vrule width 0pt height 15pt depth 5pt}|c|c|c|c|}
    \hline\hline
    \WOPPER/   & \.{HERWIG}   & semantics   & Default value
    \\\hline\hline
    \.{hwqcdl} & \.{QCDLAM}   & $\Lambda_{QCD}/\mathop{\rm GeV}$
                                            & 0.18
    \\\hline
    \.{hwrms1} & \.{RMASS(1)} & $m_d/\mathop{\rm GeV}$
                                            & 0.32
    \\\hline
    \.{hwrms2} & \.{RMASS(2)} & $m_u/\mathop{\rm GeV}$
                                            & 0.32
    \\\hline
    \.{hwrms3} & \.{RMASS(3)} & $m_s/\mathop{\rm GeV}$
                                            & 0.5
    \\\hline
    \.{hwrms4} & \.{RMASS(4)} & $m_c/\mathop{\rm GeV}$
                                            & 1.8
    \\\hline
    \.{hwrms5} & \.{RMASS(5)} & $m_b/\mathop{\rm GeV}$
                                            & 5.2
    \\\hline
    \.{hwrms6} & \.{RMASS(6)} & $m_t/\mathop{\rm GeV}$
                                            & 100.00
    \\\hline
    \.{hwrms0} & \.{RMASS(13)} & $m_g^{eff.}/\mathop{\rm GeV}$
                                            & 0.75
    \\\hline
    \.{hwvqcu} & \.{VQCUT}    & Quark virtuality cutoff
                                            & 0.48
    \\\hline
    \.{hwvgcu} & \.{VGCUT}    & Gluon virtuality cutoff
                                            & 0.10
    \\\hline
    \.{hwvpcu} & \.{VPCUT}    & Photon virtuality cutoff
                                            & -1.00%
       \footnote{Negative values will translate to $\sqrt{s}$.}
    \\\hline
    \.{hwclma} & \.{CLMAX}    & Max.~cluster mass parameter
                                            & 3.35
    \\\hline
    \.{hwpspl} & \.{PSPLT}    & Split cluster parameter
                                            & 1.00
    \\\hline
    \.{hwqdiq} & \.{QDIQK}    & Max.~scale for $g\to\mathop{\rm diquark}$
                                            & 0.00
    \\\hline
    \.{hwpdiq} & \.{PDIQK}    & $g\to\mathop{\rm diquark}$ rate parameter
                                            & 5.00
    \\\hline
    \.{hwqspa} & \.{QSPAC}    & Spacelike evolution cutoff
                                            & 2.50
    \\\hline
    \.{hwptrm} & \.{PTRMS}    & Intrinsic $p_T$
                                            & 0.00
    \\\hline
  \end{tabular}
  \end{center}
  \end{minipage}
  \caption{\.{HERWIG} parameters accessible from \WOPPER/.}
  \label{tab:wopper-herwig-parm}
\end{table}

\begin{table}
  \begin{minipage}{\textwidth}
  \begin{center}
  \begin{tabular}{@{\vrule width 0pt height 15pt depth 5pt}|c|c|c|c|}
    \hline\hline
    \WOPPER/   & \.{HERWIG}   & semantics   & Default value
    \\\hline\hline
    \.{hwipri} & \.{IPRINT}   & Printout option
                                            & 1
    \\\hline
    \.{hwmaxp} & \.{MAXPR}    & Max. printouts
                                            & 0
    \\\hline
    \.{hwmaxe} & \.{MAXER}    & Max. errors
                                            & 10
    \\\hline
    \.{hwlwev} & \.{LWEVT}    & Event output unit
                                            & 0
    \\\hline
    \.{hwlrsu} & \.{LRSUD}    & Sudakov input unit
                                            & 0
    \\\hline
    \.{hwlwsu} & \.{LWSUD}    & Sudakov output unit
                                            & 77
    \\\hline
    \.{hwsudo} & \.{SUDORD}   & Sudakov order in $\alpha_S$
                                            & 1
    \\\hline
    \.{hwnrn1} & \.{NRN(1)}   & First random seed
                                            & 17673
    \\\hline
    \.{hwnrn2} & \.{NRN(2)}   & Second random seed
                                            & 63565
    \\\hline
    \.{hwazso} & \.{AZSOFT}   & Soft gluon azimuthal correlations
                                            & .true.
    \\\hline
    \.{hwazsp} & \.{AZSPIN}   & Gluon spin azimuthal correlations
                                            & .true.
    \\\hline
    \.{hwb1li} & \.{B1LIM}   & $B$-cluster $\to$ 1 hadron parameter
                                            & 0.0
    \\\hline
  \end{tabular}
  \end{center}
  \end{minipage}
  \caption{Technical \.{HERWIG} parameters accessible from \WOPPER/.}
  \label{tab:wopper-technical-herwig-parm}
\end{table}


\section{\f77/ Interface}
\label{sec:f77}

\WOPPER/ version \Version/ provides two application program interfaces
on different levels.  The higher (preferred) level consists of the
command interpreter \.{wwdcmd} that accepts commands in form of
\.{character*(*)} strings.  This driver communicates with the analyzer
\hepawk/ \cite{Ohl92a} by default.  The lower level consists of two \f77/
subroutine calls: \.{wwpsrv} and \.{wopper}.

\subsection{Higher Level Interface}
\label{sec:f77-high-level}

\begin{example}{Higher level \f77/ interface}{ex:f77}
\.{* wopperappl.f}                                                \\
\>\>\> \.{...}                                                    \\
\>\>\> \.{call wwdcmd ('init')}   \C initialize the generator     \\
\>\>\> \.{...}                                                    \\
\>\>\> \.{call wwdcmd ('generate 10000')}
                                   \C generate 10000 events       \\
\>\>\> \.{...}                                                    \\
\>\>\> \.{call wwdcmd ('close')}  \C cleanup                      \\
\>\>\> \.{...}
\end{example}

The simple commands understood by \.{wwdcmd} are  (here keywords are
typeset in typewriter font and variables in italics; vertical bars
denote alternatives)
\begin{itemize}
  \item \.{initialize} \hfil\goodbreak
    Force initialization of \WOPPER/ and write an initialization
    record into the
    \hepevt/ event record, which should trigger the necessary
    initializations in the analyzer.
  \item \.{generate} $[n]$ \hfil\goodbreak
    Generate \.{nevent} events and call \hepawk/ to analyze them.
    If the optional parameter $n$ is supplied, \.{nevent} is set
    to its value.
  \item \.{close} \hfil\goodbreak
    Write a termination record to \hepevt/, which should
    trigger the necessary cleanups in the analyzer.
  \item \.{statistics} \hfill\goodbreak
    Print performance statistics (this is usually only useful for the
    \WOPPER/ developers, who are tuning internal parameters).
  \item \.{quit} \hfil\goodbreak
    Terminate \WOPPER/ without writing a termination
    record.
  \item \.{exit}$\vert$\.{bye} \hfil\goodbreak
    Write a termination record and terminate \WOPPER/.
  \item \.{set} {\it variable\/}
         {\it ival\/}$\vert${\it rval\/} \hfil\goodbreak
    Set physical or internal parameters.  See the tables
    \ref{tab:wopper-physics-parm} and \ref{tab:wopper-technical-parm}
    for a comprehensive listing of all variables.  For example, the
    command \verb+'set ahpla 127.0'+ will set the QED fine structure
    constant to $1/127$.
  \item \.{print} {\it variable\/}$\vert$\.{all} \hfil\goodbreak
    Print the value of physical or internal variables.
    Specifying the special variable \.{all} causes a listing
    of all variables known to \WOPPER/.
  \item \.{debug}$\vert$\.{nodebug} {\it flag\/} \hfil\goodbreak
    Toggle debugging flags.
  \item \.{testran} \hfil\goodbreak
    Test the portability of the random number generator.  We use
    a generator of the Marsaglia-Zaman variety \cite{MZT90},
    which should give
    identical results on almost all machines.
  \item \.{banner} \hfil\goodbreak
    Print a string identifying this version of \WOPPER/.
  \item \.{echo} {\it message\/} \hfil\goodbreak
    Print {\it message\/} on standard output.
  \item \.{lugive} {\it string\/} \hfil\goodbreak
    Pass {\it string\/} to {\tt JETSET}'s {\tt LUGIVE()} routine.
    See~\cite{JETSET} for details.
\end{itemize}

Unique abbreviations of the keywords are accepted, i.e.~\verb+'g 1000'+
generates 1000 events.  The tokens are separated by blanks. Blank
lines and lines starting with a \.{\#} are ignored and may be used for
comments.  For portability, only the first 72 characters of each line
are considered.

On UNIX systems \WOPPER/ reads the default startup files {\tt .wopper}
in the user's home directory and the current directory, if they exist.


\subsection{Lower Level Interface}
\label{sec:f77-low-level}

The subroutine \.{wopper (code)} has a single integer parameter.
The parameter \.{code} is interpreted as follows:
\begin{itemize}
  \item{} 0: initialize the generator and write an initialization
      record to \hepevt/.
  \item{} 1: generate an event and store it in \hepevt/.  If \WOPPER/
    has not been initialized yet, the necessary initializations are
    performed, but no initialization record is written.
  \item{} 2: perform final calculations and write the results
      to \hepevt/.
\end{itemize}
Figure \ref{ex:f77-wwdcmd} displays excerpts from a simplified
version of \.{wwdcmd} that make the correspondences between the
two levels of the \f77/ interface explicit.

\begin{example}{Event generation loop}{ex:f77-wwdcmd}
\.{* wwdcmd.f}                                                    \\
\>\>\> \.{subroutine wwdcmd (cmdlin)}                             \\
\>\>\> \.{character*(*) cmdlin}                                   \\
\>\>\> \.{...}                                                    \\
\>\>\> \.{else if (cmdlin.eq.'init')}                             \\
\>\>\> \>\> \.{call wopper (0)}        \C initialize \WOPPER/     \\
\>\>\> \>\> \.{call hepawk ('scan')}   \C print
                                            initialization record \\
\>\>\> \.{else if (cmdlin.eq.'generate')}                         \\
\>\>\> \>\> \.{do 10 n = 1, nevent}                               \\
\>\>\> \>\> \>\> \.{call wopper (1)}      \C generate an event    \\
\>\>\> \>\> \>\> \.{call hepawk ('scan')} \C analyze the event    \\
\.{10} \>\>\>\>\> \.{continue}                                    \\
\>\>\> \.{else if (cmdlin.eq.'close')}                            \\
\>\>\> \>\> \.{call wopper (2)}        \C get total cross
                                                    section, etc. \\
\>\>\> \>\> \.{call hepawk ('scan')}   \C finalize analysis       \\
\>\>\> \.{else}                                                   \\
\>\>\> \.{...}                                                    \\
\>\>\> \.{end}
\end{example}

\WOPPER/'s parameters can be accessed on the lower level by the
subroutine \.{wwpsrv (result, action, name, type, ival, rval, dval,
  lval)}. The parameter is specified by its (lowercase) name in the
\.{character*(*)} string \.{name}.  The string \.{action} is either
\verb+'read'+ or \verb+'write'+ corresponding to whether the parameter
is to be inspected or modified. The type of the parameter (\verb+'int'+,
\verb+'real'+, \verb+'dble'+, or \verb+'lgcl'+) is specified in
\.{type}; it is an input parameter for \verb+'read'+ and an output
parameter for \verb+'write'+.  Depending on this type the value is
passed in \.{ival}, \.{rval}, \.{dval}, or \.{lval}, respectively.  The
following error codes will be returned in the string \.{result}:
\verb+' '+: no error, \verb+'enoarg'+: invalid \.{action},
\verb+'enoent'+: no such parameter, \verb+'enoperm'+: permission
denied, and \verb+'enotype'+: invalid \.{type}.

The protection scheme implemented with this parameter handling has
been described in \cite{ADM+92a}.  Its main purpose is to guarantee
consistency of user defined and computed parameters in the generation
phase of the Monte Carlo.

\section{Conclusions}
\label{sec:concl}

We have presented version \Version/ of the Monte Carlo event generator
\WOPPER/ for $W$ pair production and decays into four fermions at high
energy $e^+e^-$ colliders.  The distinguishing feature of \WOPPER/ is
the inclusion of higher order electromagnetic corrections including soft
photon exponentiation and explicit generation of exclusive hard photons.
In contrast to fixed order calculations which have to be exponentiated
by hand, \WOPPER/ handles the multiphoton effects explicitly.

The present version does not contain weak corrections.  Forthcoming
versions of the Monte Carlo generator will include weak corrections in
the framework of effective Born cross sections \cite{DBD92:FKK+92}.
Anomalous couplings for the
$\gamma WW$ and $ZWW$ vertices may also be included in a future version
of the generator.

\section*{Acknowledgments}
\label{sec:ack}

We are grateful to
\begin{itemize}
  \item{} Hywel Phillips of DELPHI for help with the {\tt JETSET}
    interface and for tireless bug hunting.
  \item{} Tj\"orb\"orn Sj\"ostrand (CERN/TH) for {\tt JETSET} and for
    supporting it.
  \item{} Frits Erne of L3 for pointing out a bug to us.
  \item{} The members of the $W$-physics working group of the 1995
    LEP2 workshop for appreciating our efforts.
\end{itemize}

\appendix
\section{Distribution}

The latest release of \WOPPER/ is available by anonymous ftp from
\begin{verbatim}
  crunch.ikp.physik.th-darmstadt.de
\end{verbatim}
in the directory
\begin{verbatim}
  pub/ohl/wopper
\end{verbatim}
or on the World Wide Web at the URL
\begin{verbatim}
  http://crunch.ikp.physik.th-darmstadt.de/monte-carlos.html#wopper
\end{verbatim}
Important announcements (new versions, fatal bugs, etc.)
will be made through the mailing list
\begin{verbatim}
  wopper-announce@crunch.ikp.physik.th-darmstadt.de
\end{verbatim}
Subscriptions can be obtained from
\begin{verbatim}
  majordomo@crunch.ikp.physik.th-darmstadt.de
\end{verbatim}
(send a message consisting of \texttt{help} to \texttt{majordomo} for
instructions on how to subscribe, don't send such messages to the list
itself).

\section{Installation}

\WOPPER/ is distributed in PATCHY format \cite{KZ88}.
Plain \f77/ versions can be made available on request.  Since the \f77/
source conforms to ANSI X3.9-1978, it should run without modifications
on all platforms.

\subsection{UNIX Systems}

On UNIX systems, the configuration, compilation and installation
can be performed automatically according to the following familiar
sequence:
\begin{verbatim}
    $ ./configure
    $ make
    $ make test
    $ make install
\end{verbatim}

Figure~\ref{fig:configure} shows the command line options of the
\texttt{configure} script for \WOPPER/ on UNIX systems.  This
\texttt{configure} script has been created by the popular GNU
Autoconf\cite{autoconf} package and should work on all UNIX variants.

\begin{figure}
{\small\begin{verbatim}
Usage: configure [options] [host]
Options: [defaults in brackets after descriptions]
Configuration:
  --cache-file=FILE       cache test results in FILE
  --help                  print this message
  --no-create             do not create output files
  --quiet, --silent       do not print `checking...' messages
  --version               print the version of autoconf that created configure
Directory and file names:
  --prefix=PREFIX         install architecture-independent files in PREFIX
                          [/usr/local]
  --exec-prefix=PREFIX    install architecture-dependent files in PREFIX
                          [same as prefix]
  --srcdir=DIR            find the sources in DIR [configure dir or ..]
  --program-prefix=PREFIX prepend PREFIX to installed program names
  --program-suffix=SUFFIX append SUFFIX to installed program names
  --program-transform-name=PROGRAM run sed PROGRAM on installed program names
Host type:
  --build=BUILD           configure for building on BUILD [BUILD=HOST]
  --host=HOST             configure for HOST [guessed]
  --target=TARGET         configure for TARGET [TARGET=HOST]
Features and packages:
  --disable-FEATURE       do not include FEATURE (same as --enable-FEATURE=no)
  --enable-FEATURE[=ARG]  include FEATURE [ARG=yes]
  --with-PACKAGE[=ARG]    use PACKAGE [ARG=yes]
  --without-PACKAGE       do not use PACKAGE (same as --with-PACKAGE=no)
  --x-includes=DIR        X include files are in DIR
  --x-libraries=DIR       X library files are in DIR
--enable and --with options recognized:
  --with-g77              use GNU Fortran 77
  --enable-verbose-patchy display all patchy output
  --enable-internal       do not use this!
  --enable-notime         do not use timing functions
  --enable-pedantic       no IMPLICIT NONE
  --with-libpath=PATH     use PATH for libraries
  --with-srcpath=PATH     use PATH for source files (CARs)
  --with-hepawk           use HEPAWK
  --with-jetset           use JETSET
  --with-herwig=CAR       use HERWIG CAR file
  --with-cernlib          use CERNLIB
  --enable-paper-a4       use European (A4) paper
  --enable-paper-us       use US (letter) paper
\end{verbatim}}
\caption{\label{fig:configure}%
  Comandline options of the \texttt{configure} script for
  \WOPPER/ on UNIX systems.}
\end{figure}

\subsection{Non-UNIX Systems}

For non-UNIX systems configuration and compilation has to be performed
manually from the \texttt{CARDS} file and the \texttt{cradle}s.

\section{External Symbols:
         Common Blocks and Subroutines}
\label{sec:ext-names}

To avoid possible name clashes with other packages, all external symbols
exported by \WOPPER/ begin with the two letters \.{WW}, except
for the routine \.{wopper} itself and the \hepevt/ common block.


\begin{itemize}
  \item Common Blocks: \hfil\goodbreak
    The following common blocks are used by \WOPPER/.
    \begin{itemize}
    \item \.{/hepevt/, /hepspn/}: standard common blocks for passing
      generated events \cite{AKV89}.
    \item \.{/wwpcom/}: main parameter common block, holds all physical
        parameters.  Application programs should access this common block
        through the \.{wwpsrv} routine.
    \item \.{/wwcbrn/}: internal parameters used for born cross section.
    \item \.{/wwcdec/}: internal parameters used for $W$ decays.
    \item \.{/wwcevt/}: internal parameters used for event generation.
    \item \.{/wwcsta/}: statistics.
    \item \.{/wwctri/}: storage for keyword lookup.
    \end{itemize}
  \item Driver Program: \hfil\goodbreak
    \begin{itemize}
    \item \.{wwdriv}: sample main program, which reads commands
      from standard input and feeds them into \.{wwdcmd}.
    \item \.{wwdloo}: command loop, reading command from a terminal or
      file and executing them.
    \item \.{wopper}: the low level entry point into the generator
      for application programs.
    \item \.{wwdcmd}: \WOPPER/'s command interpreter, the preferred
      entry point for application programs.  Executes a single command.
    \item \.{wwdlxi, wwdlxd, wwdlxs}: Utility routines: tokenization
      of input.
    \item \.{wwdsig}: UNIX signal handler.
    \item \.{isatty}: check if this job is run interactively.
  \end{itemize}
  \item Parameter Management: \hfil\goodbreak
    These routines are used to control the parameters common block
    \.{/wwpcom/}.
  \begin{itemize}
    \item \.{wwpsrv}: server handling parameter changing requests.
    \item \.{wwpini}: \.{block data} supplying default values.
    \item \.{wwpprn}: print parameters.
  \end{itemize}
  \item Initialization:\hfil\goodbreak
  \begin{itemize}
    \item \.{wwinit}: main entry point for initializations.
    \item \.{wwigsw}: initialization of electroweak parameters.
    \item \.{wwicut}: initialization of internal Monte Carlo parameters.
    \item \.{wwibmx}: finds maximum of Born cross section.
    \item \.{wwibn}: auxiliary function for finding maximum of on-shell
      cross section.
    \item \.{wwibns}: auxiliary function for finding maximum of
      cross section for symmetric virtualities.
    \item \.{wwibna}: auxiliary function for finding maximum of
      cross section for one $W$ on-shell.
    \item \.{wwibnv}:
  \end{itemize}
  \item Final calculations:
  \begin{itemize}
    \item \.{wwclos}: calculate total cross section, errors and close
      the generator.
    \item \.{wwstat}: statistics.
  \end{itemize}
  \item Hard Subprocess Generation:\hfil\goodbreak
  \begin{itemize}
    \item \.{wwgen}: main entry point for hard subprocess generation.
    \item \.{wwgt}: generation of angular distribution of virtual $W$'s.
    \item \.{wwgppr}: generate four-momenta of virtual $W$'s.
    \item \.{wwgdec}: generate final state fermions from $W$ decays.
    \item \.{wwdqfl}: select quark flavors in W decay.
  \end{itemize}
  \item Branching: \hfil\goodbreak
  \begin{itemize}
    \item \.{wwbini}: generates the initial state photon radiation.
  \end{itemize}
  \item Cross Sections: \hfil\goodbreak
  \begin{itemize}
    \item \.{wwxhel}: helicity amplitudes for $W$ pair production.
    \item \.{wwxtot}: total off-shell cross section.
    \item \.{wwxint}: auxiliary function for integration over virtual
      $W$ masses.
  \end{itemize}
  \item Accessing \hepevt/: \hfil\goodbreak
  \begin{itemize}
    \item \.{wweeni}: enter identification of Monte Carlo and run.
    \item \.{wweens}: write summary record
    \item \.{wweent}: enter one particle
    \item \.{wwenul}: enter null particle
    \item \.{wwenew}: clear \hepevt/
  \end{itemize}
  \item Hadronization: \hfil\goodbreak
  \begin{itemize}
    \item \.{wwpart}: leave partons alone.
    \item \.{wwlund}: {\tt JETSET} interface code
    \item \.{wwhwig}: {\tt HERWIG} interface code
    \item \.{hwaend}: {\tt HERWIG} abnormal end routine
  \end{itemize}
  \item Random Numbers: \hfil\goodbreak
  \begin{itemize}
    \item \.{wwrgen}: returns a double precision uniform deviate.
    \item \.{wwrmz}: random number generator \.{RANMAR}.
    \item \.{wwrlux}: random number generator \.{RANLUX}.
    \item \.{wwrtst}: test the portability of the random number
                        generator.
    \item \.{wwrtmz}: test \.{RANMAR}.
    \item \.{wwrtlx}: test \.{RANLUX}.
    \item \.{wwrchk}: used in testing \.{RANLUX}.
    \item \.{rluxat, rluxgo, rluxin, rluxut}: additional \.{RANLUX}
      entry points which are not used by \WOPPER/.
  \end{itemize}
  \item Utilities: \hfil\goodbreak
  \begin{itemize}
    \item \.{wwumsg}: messages and error exit.
    \item \.{wwulwr}: convert input to lower case.
    \item \.{wwuboo}: boost a four vector.
    \item \.{wwutim}: still available CPU time for this job.
    \item \.{wwuamo}: multidimensional minimization.
    \item \.{wwumin}: onedimensional minimization.
    \item \.{wwuons}: Gram-Schmidt procedure.
    \item \.{wwuort}: another Gram-Schmidt procedure.
  \end{itemize}
  \item Canonical cuts: \hfil\goodbreak
  \begin{itemize}
    \item \.{adloth}: apply canonical cuts in \hepevt/.
    \item \.{adloip}: inner product of vectors in \hepevt/.
    \item \.{adloan}: angle to beam in \hepevt/.
    \item \.{adload}: add four momenta in \hepevt/.
  \end{itemize}
  \item Keyword search: \hfil\goodbreak
    (using the dynamic tries described in \cite{Dun91}).
  \begin{itemize}
    \item \.{wwtins}: insert a new keyword.
    \item \.{wwtlup}: look up a (possibly abbreviated) keyword.
    \item \.{wwtnew}: insert new a node into the trie.
    \item \.{wwtlen}: calculate length of keyword.
    \item \.{wwtc2a}: convert keyword from \.{character*(*)} to
      \.{integer(*)}.
  \end{itemize}
\end{itemize}



\newpage

\section*{Test Run}

\WOPPER/ version \Version/ is distributed together with a sample command
file and \hepawk/ script, which are given below.  To run this example,
the user will need to link \WOPPER/ with the CERN library, because
histogramming is done by HBOOK \cite{HBOOK}.

The file {\tt sample.wopper} is read from standard input ({\tt unit
  stdin}, which is initialized to 5), and {\tt sample.hepawk} is read
from the file {\tt SCRIPT} (i.e.\ under MVS from the file which has been
allocated to the {\tt DDNAME SCRIPT} and under UNIX from the file {\tt
  script} or from the value of the environment variable {\tt SCRIPT}).


\subsection*{{\tt sample.wopper}}

Here is a simple \WOPPER/ command file, setting up parameters and
generating 10000 events.

{\small
\begin{verbatim}
# sample.wopper -- sample WOPPER command file

# parameters
set ebeam  87.5

# run
init
gen 10000
close
quit
\end{verbatim}
} 


\subsection*{{\tt sample.hepawk}}

This is a small \hepawk/ script that counts the muons from the $W$
decays and plots a histogram of their energy distribution.  The first
generated event is dumped to illustrate the usage of the \hepevt/
common block.

{\small
\begin{verbatim}
# sample.hepawk -- sample HEPAWK analyzer for WOPPER.

BEGIN
  {
    printf ("\nWelcome to the WOPPER test:\n");
    printf ("***************************\n\n");

    printf ("Monte Carlo Version: %s\n", REV);
    printf ("                Run: %d\n", RUN);
    printf ("               Date: %s\n\n", DATE);

    E_max = 100;
    N_chan = 50;

    h_muon_energy
      = book1 (0, "Muon-energy", N_chan, 0, E_beam);
    nr_muons = 0;               # initialize counter
    dumped_an_event = 0;
  }

  {
    if (dumped_an_event == 0)
      {
        dump ("vs");            # Dumping first event
        dumped_an_event++;
      }

    for (@p in LEPTONS)
      {
      if (abs(@p:id) == _pdg_muon)
        {
          fill (h_muon_energy, @p:p:E);
          nr_muons++;
        }
      }
  }

END
  {
    # Dump some numbers
    printf ("\nRESULTS:\n");
    printf ("********\n\n");

    printf ("Total events:    %d, total cross section: %g pb\n",
            NEVENT, XSECT * 1e9);
    printf ("Number of muons: %d\n\n", nr_muons);

    plot();
    printf ("\ndone.\n");
  }
\end{verbatim}
} 


\newpage

\subsection*{{\tt sample.output}}

The following output should result from the input files above, up to
small roundoff errors and different \f77/ default output formats.

{\small
\begin{verbatim}
wwdcmd: message: Starting WOPPER, Version 1.02/99, (build      0/0000)
hepawk: message: starting HEPAWK, Version 1.6
wwigsw: message: *******************************************************
wwigsw: message: "GF scheme" selected:
wwigsw: message: Using GFERMI and ALPHA as input, calculating SIN2W.
wwigsw: message: Using derived W width.
wwigsw: message: Parameters used in this run:
wwigsw: message: AHPLA  =    128.00000 ( = 1/alpha(2 M_W) )
wwigsw: message: SIN2W  =       .23121 (effective mixing angle)
wwigsw: message: GFERMI =  .116639E-04 GeV**(-2)
wwigsw: message: ALPHAS =       .12000
wwigsw: message: GAMM1W =      2.08468 GeV (S.M. value, used)
wwigsw: message: CKMVUS =       .21960 CKM Matrix (Cabibbo angle)
wwigsw: message: CKMVCB =       .04000 CKM Matrix
wwigsw: message: CKMVUB =       .00320 CKM Matrix
wwigsw: message: Z-e-e couplings:
wwigsw: message:    g_V =      -.01396
wwigsw: message:    g_A =      -.18579
wwigsw: message: W-e-nu coupling:
wwigsw: message:      g =       .23038
wwigsw: message: Z-W-W coupling:
wwigsw: message:  g_ZWW =       .57134
wwigsw: message: gamma-W-W coupling:
wwigsw: message:  g_gWW =       .31333
wwigsw: message: Using energy-dependent width for W and Z propagators.
wwigsw: message: *******************************************************

Welcome to the WOPPER test:
***************************

Monte Carlo Version: v01.02 (*** 00 00:00:00  1900)
                Run: 1035996352
               Date: Apr 10 03:40:00  1995

========================================================================
Dumping the event record for event #         1
There are   13 entries in this record:
------------------------------------------------------------------------
Entry #   1 is an incoming (HERWIG convention) electron
p: ( .8750E+02;  .0000E+00,  .0000E+00,  .8750E+02),  m:  .0000E+00
------------------------------------------------------------------------
Entry #   2 is an incoming (HERWIG convention) positron
p: ( .8750E+02;  .0000E+00,  .0000E+00, -.8750E+02),  m:  .0000E+00
------------------------------------------------------------------------
Entry #   3 is the CMS system (HERWIG convention)
p: ( .1750E+03;  .0000E+00,  .0000E+00,  .0000E+00),  m:  .1750E+03
------------------------------------------------------------------------
Entry #   4 is a null entry.
------------------------------------------------------------------------
Entry #   5 is a null entry.
------------------------------------------------------------------------
Entry #   6 is a null entry.
------------------------------------------------------------------------
Entry #   7 is reserved for model builders.
------------------------------------------------------------------------
Entry #   8 is reserved for model builders.
------------------------------------------------------------------------
Entry #   9 is an existing photon
p: ( .1554E+00;  .6972E-03, -.1679E-03, -.1554E+00),  m:  .0000E+00
The mother is the positron                  #   5.
------------------------------------------------------------------------
Entry #  10 is an existing electron
p: ( .5965E+02;  .5462E+02,  .1940E+02,  .1411E+02),  m:  .5110E-03
The first mother is the W-                        #   7.
The other mother is the anti-electron-neutrino    #  11.
------------------------------------------------------------------------
Entry #  11 is an existing anti-electron-neutrino
p: ( .2912E+02; -.2804E+02, -.3085E+01, -.7203E+01),  m:  .0000E+00
The first mother is the W-                        #   7.
The other mother is the electron                  #  10.
------------------------------------------------------------------------
Entry #  12 is an existing up-quark
p: ( .3692E+02;  .1710E+02, -.3106E+02,  .1030E+02),  m:  .5000E-02
The first mother is the W+                        #   8.
The other mother is the anti-down-quark           #  13.
------------------------------------------------------------------------
Entry #  13 is an existing anti-down-quark
p: ( .4915E+02; -.4368E+02,  .1474E+02, -.1705E+02),  m:  .1000E-01
The first mother is the W+                        #   8.
The other mother is the up-quark                  #  12.
========================================================================

RESULTS:
********

Total events:         10000, total cross section:  12.98     pb
Number of muons:       2194

1Muon-energy

 HBOOK     ID =         1                                        DATE  10/04/95

      160
      156                               7 X
      152                               XXX
      148                               XXX
      144                             2 XXX5
      140                             X XXXX
      136                             X2XXXX72
      132                           2 XXXXXXXX
      128                           X XXXXXXXX
      124                           XXXXXXXXXX
      120                          7XXXXXXXXXX
      116                          XXXXXXXXXXX
      112                          XXXXXXXXXXX
      108                          XXXXXXXXXXX
      104                        5 XXXXXXXXXXX
      100                      X7X XXXXXXXXXXX
       96                      XXX XXXXXXXXXXX
       92                      XXX2XXXXXXXXXXX
       88                      XXXXXXXXXXXXXXX
       84                      XXXXXXXXXXXXXXX
       80                      XXXXXXXXXXXXXXX7
       76                      XXXXXXXXXXXXXXXX
       72                     7XXXXXXXXXXXXXXXX
       68                     XXXXXXXXXXXXXXXXX
       64                     XXXXXXXXXXXXXXXXX
       60                     XXXXXXXXXXXXXXXXX
       56                     XXXXXXXXXXXXXXXXX
       52                    2XXXXXXXXXXXXXXXXX
       48                    XXXXXXXXXXXXXXXXXX
       44                    XXXXXXXXXXXXXXXXXX
       40                    XXXXXXXXXXXXXXXXXX
       36                    XXXXXXXXXXXXXXXXXX
       32                    XXXXXXXXXXXXXXXXXX
       28                    XXXXXXXXXXXXXXXXXX5
       24                   7XXXXXXXXXXXXXXXXXXX
       20                   XXXXXXXXXXXXXXXXXXXX
       16                   XXXXXXXXXXXXXXXXXXXX2
       12                   XXXXXXXXXXXXXXXXXXXXX
        8                   XXXXXXXXXXXXXXXXXXXXXX
        4                 7XXXXXXXXXXXXXXXXXXXXXXX57 7  2

 CHANNELS  10   0        1         2         3         4         5
            1   12345678901234567890123456789012345678901234567890

 CONTENTS 100                  1 1 11111111111
           10               247090812243555433721
            1.            34391092999413526253963823 3  1

 LOW-EDGE  10        111112222233333444445555566666777778888899999
            1.   2468024680246802468024680246802468024680246802468

 * ENTRIES =       2194      * ALL CHANNELS =  .2194E+04      * UNDERFLOW =
.00
 * BIN WID =  .2000E+01      * MEAN VALUE   =  .4588E+02      * R . M . S =
.10

done.
wwdriv: message: bye.
\end{verbatim}
} 


\section{Revision History}
\label{sec:history}

\subsection*{Version 1.3, April 1995}
\begin{itemize}
  \item{} Canonical cuts and input parameters.
  \item{} Fixed inconsistent phase conventions, which resulted in
    wrong angular distributions.
\end{itemize}

\subsection*{Version 1.2, July 1994}
\begin{itemize}
  \item{} Coulomb correction.
  \item{} Improved \.{JETSET} and \.{HERWIG} interfaces.
\end{itemize}

\subsection*{Version 1.1, February 1994}
\begin{itemize}
  \item{} Hadronization, \.{JETSET} and \.{HERWIG} interfaces.
  \item{} Minor bug fixes.
\end{itemize}

\subsection*{Version 1.0, 1993}

First official release, submitted to the {\em Computer Physics
Communication Library}.

\end{document}